\documentclass[12pt,preprint]{aastex}
\usepackage{amsmath,natbib,psfig,emulateapj5}

\input{buote.defs}

\begin{document} 

\title{The Dark Matter Radial Profile in the Core of the Relaxed Cluster A2589}
\author{David A. Buote and Aaron D. Lewis} 
\affil{Department of Physics and Astronomy, University of
California at Irvine, 4129 Frederick Reines Hall, Irvine, CA
92697-4575, buote@uci.edu, lewisa@uci.edu}

\slugcomment{To Appear in The Astrophysical Journal March 20 Issue}

\begin{abstract}  

We present an analysis of a \chandra--ACIS observation of the galaxy
cluster A2589 to constrain the radial distribution of the total
gravitating matter and the dark matter in the core of the
cluster. A2589 is especially well-suited for this analysis because the
hot gas in its core region ($r\la 0.1\rvir$) is undisturbed by
interactions with a central radio source. From the largest radius
probed ($r=0.07\rvir$) down to $r\approx 0.02\rvir$ dark matter
dominates the gravitating mass. Over this region the radial profiles
of the gravitating and dark matter are fitted well by the NFW and
Hernquist profiles predicted by CDM. The density profiles are also
described well by power laws, $\rho\propto r^{-\alpha}$, where
$\alpha=1.37\pm 0.14$ for the gravitating matter and $\alpha=1.35\pm
0.21$ for the dark matter. These values are consistent with profiles
of CDM halos but are significantly larger than $\alpha\approx 0.5$
found in LSB galaxies and expected from self-interacting dark matter
models.

\end{abstract}

\keywords{galaxies:clusters:individual (A2589) --- dark matter ---
intergalactic medium --- X-ray:galaxies:clusters --- cosmological
parameters}

\section{Introduction}
\label{intro}

The very precise constraints that have been placed recently on the
cosmological world model by observations of the cosmic microwave
background \citep[e.g.,][]{sper03a} and high-redshift supernovae
\citep[e.g.,][]{perl99a} require that most of the matter in the
universe is non-baryonic ``dark matter'' and that the largest
contributor to the energy density of the universe is the ``dark
energy''. But these two mysterious quantities that dominate the energy
density of the universe still could be merely fitting parameters --
akin to Ptolemaic epicycles -- that are not physical quantities.
There is, therefore, great urgency to discover the nature of the dark
matter and dark energy which are the foundations of the new
cosmological paradigm.

The structure of dark matter (DM) halos is a sensitive probe of the
properties of the DM. In the standard $\Lambda$CDM paradigm, N-body
simulations show that the radial density profiles of DM halos are
fairly uniform, approximately parameterized by the NFW profile,
$\rho(r)\propto r^{-1}(r_s+r)^{-2}$ \citep{nfw}. At small radii ($r\ll
r_s)$ the density profiles follow a power law, although the precise
value of the power-law exponent remains controversial; i.e., $\rho(r)
\propto r^{-\alpha}$, with $\alpha=1$ according to NFW and
$\alpha=1.5$ according to the simulations of
\citet{moor99a}. Observations of the rotation curves of low-surface
brightness (LSB) galaxies \citep[e.g.,][]{swat00a} suggest a profile
for the DM that is substantially flatter $(\alpha\approx 0.5$) than
$\Lambda$CDM in the central regions.  These observations inspired
\citet{sper00} to propose the existence of ``self-interacting'' DM
(SIDM). In the SIDM model the DM particles are assumed to possess some
cross section for elastic collisions with each other.  Detailed CDM
simulations incorporating the SIDM idea have confirmed that the DM
profiles of LSB galaxies can be flattened as observed
\citep[e.g.,][]{dave01a}.

Like LSB galaxies, galaxy clusters provide excellent venues to study
DM because they are DM-dominated from deep down into their cores
\citep[$\approx 0.01\rvir$; e.g.,][]{dubi98a} out to their virial
radii (1\rvir). Moreover, several powerful techniques are available
for probing the cluster matter distributions: optical studies of
galaxy dynamics and gravitational lensing, and X-ray observations of
the hot gas. Each of these techniques has advantages and
disadvantages. For example, some advantages of using X-ray
observations are that the hot gas in clusters traces the
three-dimensional gravitational potential with an isotropic pressure
tensor. The quality of data for the hot gas is limited only by the
sensitivity and resolution of the X-ray detectors and not by the
finite number of galaxies.

The most important limitation associated with X-ray observations of
cluster mass distributions is the assumption of hydrostatic
equilibrium. A large fraction of nearby clusters ($z<0.2$) do have
regular image morphologies and appear to be relaxed on 0.5-1 Mpc
scales \citep[e.g.,][]{mohr95a,buot96b,jone99a}.  For such clusters
with regular morphologies (i.e., not currently experiencing a major
merger) cosmological N-body simulations have determined that masses
calculated by assuming perfect hydrostatic equilibrium are generally
quite accurate \citep[e.g.,][]{tsai94a,evra96a,math99a}. But clusters
that are observed to be relaxed on 0.5-1 Mpc scales are typically
associated with cooling flows \citep[e.g.,][]{buot96b}, and \chandra\
observations have demonstrated that the inner cores of cooling flows
are highly disturbed exhibiting holes and filamentary structures
\citep[e.g.,][]{fabi00_perseus,davi01a,etto02a} which certainly raise
serious questions about the assumption of hydrostatic equilibrium.

Hence, for studies of DM it is imperative to find clusters that are
relaxed (i.e., have smooth, regular X-ray images) from deep within
their cores out to $\sim$~Mpc scales.  Our recent analysis of the
\chandra\ ACIS-S data of one such cluster
\citep[A2029,][]{lewi02a,lewi03a} has provided the strongest
constraint on the core DM profile for a galaxy cluster to date; i.e.,
$\alpha=1.19\pm 0.04$, constrained down to $\approx 0.01\rvir$.  The
core DM density profile of A2029 is consistent with the NFW profile
and rules out an important contribution from SIDM in this cluster.
Since the DM properties of A2029 may not be typical, and the expected
cosmological scatter \citep{bull01a} needs to be addressed, more X-ray
observations of clusters with undisturbed cores are very much needed.

In our present study we analyze of the core DM profile of the galaxy
cluster A2589 using a new \chandra\ ACIS-S observation. We selected
A2589 as the brightest, nearby ($z=0.0414$) cluster with a smooth
X-ray morphology \citep[according to \rosat\ images,][]{buot96b} and
without a known central radio source \citep{baue00a}.  The latter
criterion greatly increases the likelihood that the X-ray image of the
core is undisturbed. (The redshift of A2589 corresponds to an angular
diameter distance of 171~Mpc and $1\arcsec = 0.83$~kpc assuming
$H_0=70$~\kmsmpc, $\omegam=0.3$, and $\omegalambda=0.7$.)

The paper is organized as follows. In \S \ref{obs} we present the
observation and discuss the data reduction. The spectral analysis of
the ACIS-S data is discussed in \S \ref{spec}. We calculate the
gravitating matter distribution in \S \ref{grav}, the gas mass and gas
fraction in \S \ref{gas}, and estimate the stellar mass in \S
\ref{stars}. The DM profile and the systematic error budget are
discussed in \S \ref{dm} and \S \ref{sys} respectively. Finally, in \S
\ref{conc} we present our conclusions.

\section{Observation and Data Reduction}
\label{obs}

A2589 was observed with the ACIS-S CCD camera for approximately 15~ks
during AO-3 as part of the \chandra\ Guest Observer Program. The
events list was corrected for charge-transfer inefficiency according
to \citet{town02}, and only events characterized by the standard
\asca\ grades\footnote{http://cxc.harvard.edu/udocs/docs/docs.html}
were used. The exposure time after applying these procedures is
13.7~ks. The standard \ciao\footnote{http://cxc.harvard.edu/ciao/}
software (version 2.3) was used for most of the subsequent data
preparation.

Since the diffuse X-ray emission of A2589 fills the entire S3 chip, we
attempted to use the standard background
templates\footnote{http://cxc.harvard.edu/cal} to model the
background. Although there were no strong flares (i.e., ``strong''
being defined as count rates $>10$ times nominal), there was mild
flaring activity during most of the observation. Consequently, after
running the standard {\sc lc\_clean} script to clean the source events
list of flares with the same screening criteria as the background
templates, only 3~ks remain.

To salvage a large portion of the observation (8.7~ks) while obtaining
a more accurate estimate for the background than given by the standard
templates alone, we followed the procedure discussed in
\citet{buot03a}. That is, we first extracted the source spectrum
from regions near the edge of the S3 chip and subtracted from it the
spectrum of the corresponding regions of the background
template. (Following \citealt{mark03a} we renormalized the background
template by comparing the ACIS-S1 data of the source and background
observations for energies above 10~keV. This resulted in a
normalization 8\% above nominal.)  We fitted the resultant spectrum
with a two-component model consisting of a thermal component,
represented by an \apec\ thermal plasma, and a broken power-law (BPL)
model, representing the residual flaring background which is most
pronounced at high energies. For the BPL model we obtain a break
energy of 5~keV with power-law indices of 0 below the break and 1.9
above the break. We scale this model to the area of the annular
regions of interest (see below) and generate a pha correction file for
use in \xspec; i.e., for each region analyzed below we subtract
background contributions from the standard templates and a correction
pha file accounting for the excess flaring.

\section{Spectral Analysis}
\label{spec}

\begin{figure*}[t]
\centerline{\psfig{figure=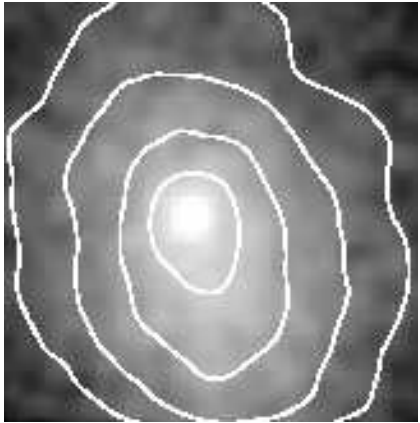,angle=0,height=0.23\textheight}}
\caption{\label{fig.image} \footnotesize 
\chandra\ ACIS-S image (0.5-5~keV) of a $4.5\arcmin \times 4.5\arcmin$
region centered on A2589. The image has been smoothed with a gaussian
filter with $\sigma=12\arcsec$. Smoothed contours overlaid are
logarithmically spaced in intensity. Note the image has not been
corrected for exposure variations, nor has the background been
subtracted. 
}
\end{figure*}

\begin{table*}[t] \footnotesize
\begin{center}
\caption{Parameters from the Spectral Fits}
\label{tab.spec}
\begin{tabular}{cllllcccc}  \tableline\tableline\\[-7pt]
& \multicolumn{2}{c}{$R_{\rm in}$} & \multicolumn{2}{c}{$R_{\rm out}$} & $T$& $\fe$ & $norm$\\
Annulus	& (arcmin) & (kpc) & (arcmin) & (kpc) & (keV) & (solar) & (10$^{-3}$~cm$^{-5}$) & $(\chi^2$/dof)\\
\tableline \\[-7pt]
1 & 0    & 0  & 0.53  & 26  & $3.16\pm 0.24$ & $1.19\pm 0.35$ & $1.22\pm 0.10$ & $53.6/56$ \\
2 & 0.53 & 26 & 0.82  & 40  & $3.14\pm 0.24$ & $0.65\pm 0.24$ & $1.41\pm 0.10$ & $68.7/57$ \\
3 & 0.82 & 40 & 1.34  & 66  & $3.30\pm 0.20$ & $0.75\pm 0.21$ & $2.72\pm 0.15$ & $118.5/104$ \\
4 & 1.34 & 66 & 1.85  & 91  & $3.03\pm 0.19$ & $0.64\pm 0.16$ & $2.86\pm 0.14$ & $128.2/107$ \\
5 & 1.85 & 91 & 3.18  & 156 & $3.18\pm 0.16$ & $0.36\pm 0.09$ & $6.74\pm 0.21$ & $165.1/167$ \\
\tableline \\[-1.0cm]
\end{tabular}
\tablecomments{Results of fitting a single \apec\ plasma emission
model modified by Galactic absorption directly to the annular spectra
(i.e., without spectral deprojection). The $norm$ parameter is the
emission measure of the \apec\ model as defined in \xspec:
$10^{-14}(\int n_en_pdV)/4\pi D^2(1+z^2)$ with units $\rm
cm^{-5}$. The quoted errors are $1\sigma$ computed using the Monte
Carlo procedure described in \S \ref{spec}.
}
\end{center}
\end{table*}

\begin{figure*}[t]
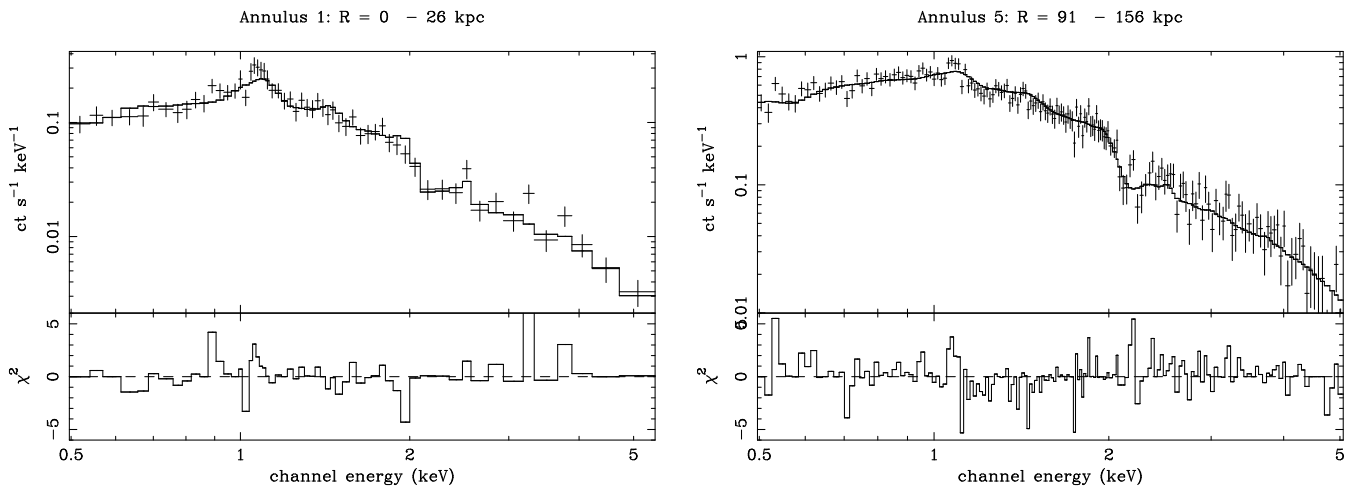

\parbox{0.49\textwidth}{
\centerline{\psfig{figure=f2a.eps,angle=-90,height=0.26\textheight}}}
\parbox{0.49\textwidth}{
\centerline{\psfig{figure=f2b.eps,angle=-90,height=0.26\textheight}
}}
\caption{\label{fig.spec} ACIS-S3 spectra accumulated within
({\sl Left Panel}) annulus \#1 and ({\sl Right Panel}) annulus
\#5. Each spectrum is fitted with an \apec\ plasma model modified by
Galactic absorption as discussed in \S \ref{spec}. }
\end{figure*} 

In Figure \ref{fig.image} we display the ACIS-S3 image of A2589. The
isophotes are regularly shaped and are approximately elliptical with
$\epsilon=0.2-0.3$ over the region of interest. The centroid of the
innermost contour is displaced with respect to that of the outer
contours shown by $\approx 15\arcsec$. Very importantly, we see no
evidence for disturbances or holes in the hot gas that are ubiquitous
in the cores of cooling flow clusters as discussed in \S
\ref{intro}. In this paper we ignore the flattening of the X-ray
isophotes and instead obtain circularly averaged spectral quantities
(i.e., gas density and temperature) on the sky to construct the
spherically averaged deprojected mass distribution.

We extracted spectra in concentric circular annuli located at the
X-ray centroid computed within a radius of $20\arcsec$ with initial
center on the peak of the X-ray emission. (We examine the sensitivity
of our results to the chosen center position in \S \ref{sys}.) The
widths of the annuli were chosen to have approximately equal
background-subtracted counts in the 0.5-5~keV band. Because of the
high background we did not use the data for $R\ga3\arcmin$, and the
widths of the annuli had to be relatively large. These restrictions
resulted in a set of five annuli within $R=3.2\arcmin$ as listed in
Table \ref{tab.spec}. We constructed ARF files for each annulus using
the latest corrections that account for the time-dependent degradation
in the quantum efficiency at lower energies. A single RMF file
appropriate for the CTI-corrected S3 data was used.

For each annulus we fitted the background-subtracted spectrum with an
\apec\ thermal plasma modified by Galactic absorption ($4.15\times
10^{20}$~\cmsq). For the \apec\ model we usually let the iron
abundance be a free parameter, and the abundances of all the other
elements are tied to iron in their solar ratios; i.e., we fit a
metallicity. The spectral fitting was performed with \xspec\
\citep{xspec} using the $\chi^2$ method. Hence, we rebinned all
spectral channels to have a minimum of 30 counts per energy bin. We
take the solar abundances in \xspec\ (v11.2.0bc) to be those given by
\citet{grsa} which use the correct ``new'' photospheric value for iron
which agrees also with the value obtained from solar-system meteorites
\citep[e.g.,][]{mcwi97}. We restricted the spectral fitting between
0.5-5~keV to avoid calibration uncertainties at lower energies and
systematic errors associated with the background subtraction at high
energies.  To estimate the statistical errors on the fitted parameters
we simulated spectra for each annulus using the best-fitting models
and fit the simulated spectra in exactly the same manner as done for
the actual data. From 100 Monte Carlo simulations we compute the
standard deviation for each free parameter which we quote as the
``$1\sigma$'' error. (We note that these $1\sigma$ error estimates
generally agree very well with those obtained using the standard
$\Delta\chi^2$ approach in \xspec.)

The parameters obtained from the spectral fits are listed in Table
\ref{tab.spec}. The simple one-temperature model is a good fit to the
data in all annuli. In Figure \ref{fig.spec} we show the ACIS-S3
spectra of the innermost and outermost annuli and the associated
best-fitting one-temperature model. Visual inspection of Figure
\ref{fig.spec} reveals that the spectral shapes of annuli \#1 and \#5
are similar above $\sim 2$~keV indicating that they have similar
temperatures. Near 1~keV the spectrum of annulus \#1 is more peaked
reflecting its larger iron abundance. In sum, the one-temperature
model reveals a nearly isothermal gas but with evidence for a
metallicity gradient similar to those observed in other clusters with
a dominant central galaxy \citep[e.g.,][]{lewi02a,etto02a,gast02a,sand02a}.

These results agree with those obtained using \rosat\ PSPC data by
\citet{davi96a} within the relatively large statistical errors
associated with the PSPC data.  \citet{davi96a} obtain a temperature
of $T=1.7^{+1.0}_{-0.3}$~keV  (90\% conf.) within $R=1\arcmin$ which is
marginally inconsistent with our results obtained with
\chandra. However, spectral parameters obtained from X-ray spectra of
groups and clusters can depend on a variety factors such as bandwidth
\citep[e.g., see][]{buot03a,buot03b}. \citet{davi96a} fitted models
over 0.3-2~keV appropriate for the PSPC. If we use 0.5-2~keV we obtain
$T\approx 2.5$~keV within $R=30\arcsec$ consistent with the PSPC
result. 

Although the fits are acceptable, we investigated whether they could
be improved further. We allowed other elemental abundances to be free
parameters. It was found that allowing silicon and oxygen to be free
did offer significant (though not substantial) improvements in the
fits in some annuli. Typically, both abundances fitted to low values
(i.e., sub-solar ratios with respect to iron) but are consistent with
the iron abundance within the estimated $1\sigma$ errors. Since it was
also found that when allowing silicon and oxygen to be free the fitted
temperatures and normalizations did not vary appreciably within their
$1\sigma$ errors, we decided to keep silicon and oxygen tied to iron
in the fits as done for the other elements.

We also found that adding another temperature component did not
improve the fits much. For example, in the central annulus adding
another temperature component improved $\chi^2$ by 3 but with the
addition of two free parameters (one for temperature and one for
normalization -- abundances are tied to the first temperature
component).  The second temperature component ($T\approx 1$~keV) has
only $\approx 1\%$ of the emission measure indicating that a single
temperature component clearly dominates the ACIS-S spectra. If we add
a high-temperature component ($T>10$~keV) to annulus \#4 the fit is
improved by $\Delta\chi^2=20$. This weak high-temperature component is
almost certainly associated with incomplete background subtraction in
this annulus. In \S \ref{sys} we discuss errors associated with the
background and other systematic effects.

\section{The Radial Profile of Gravitating Mass}
\label{grav}

\subsection{Method}
\label{method}

The approach we use to calculate the mass distribution follows closely
our previous study of A2029 \citep{lewi03a}. We approximate the hot
gas in the cluster as spherically symmetric and in hydrostatic
equilibrium so that the total gravitating mass is,
\begin{equation}
\mgrav (<r) = r{k_BT(r) \over G\mu m_{\rm p}} \left( -{d\ln
\rhog \over d\ln r} -
{d\ln T \over d\ln r} \right), \label{eqn.he}
\end{equation}
where $r$ is the radius in a three-dimensional volume, $\rho_{\rm g}$
is the gas mass density, $T$ is the gas temperature, $m_p$ is the
atomic mass unit, $\mu$ is the mean atomic weight of the gas (taken to
be 0.62), $k_B$ is Boltzmann's constant, and $G$ is Newton's
constant. $M_{\rm grav}$ is the enclosed gravitating mass which is the
sum of the masses of the stars, the gas, and the DM.  We
evaluate the derivatives of \rhog\ and $T$ using parameterized models
as discussed below.  We note that spherical averaging of the data is
appropriate for comparison to the spherically averaged DM profiles
obtained from cosmological simulations which is a key goal of this
paper.

We note that the cooling time of the gas is $\approx 1$~Gyr within the
central annulus, indicating that some heat source is required to
prevent the gas from cooling. Models of cluster cooling flows
including both feedback from an AGN and thermal conduction appear to
successfully prevent the gas from cooling
\citep[e.g.,][]{brig03a}. In the context of this paradigm, since no
substantial morphological disturbances in the X-ray image are
currently observed in the core of A2589, this cluster presumably has
had sufficient time to relax since the previous feedback episode. 

\subsection{Deprojected Radial Profiles of Gas
Density and Temperature}
\label{data}

\begin{figure*}[ht]
\parbox{0.49\textwidth}{
\centerline{\psfig{file=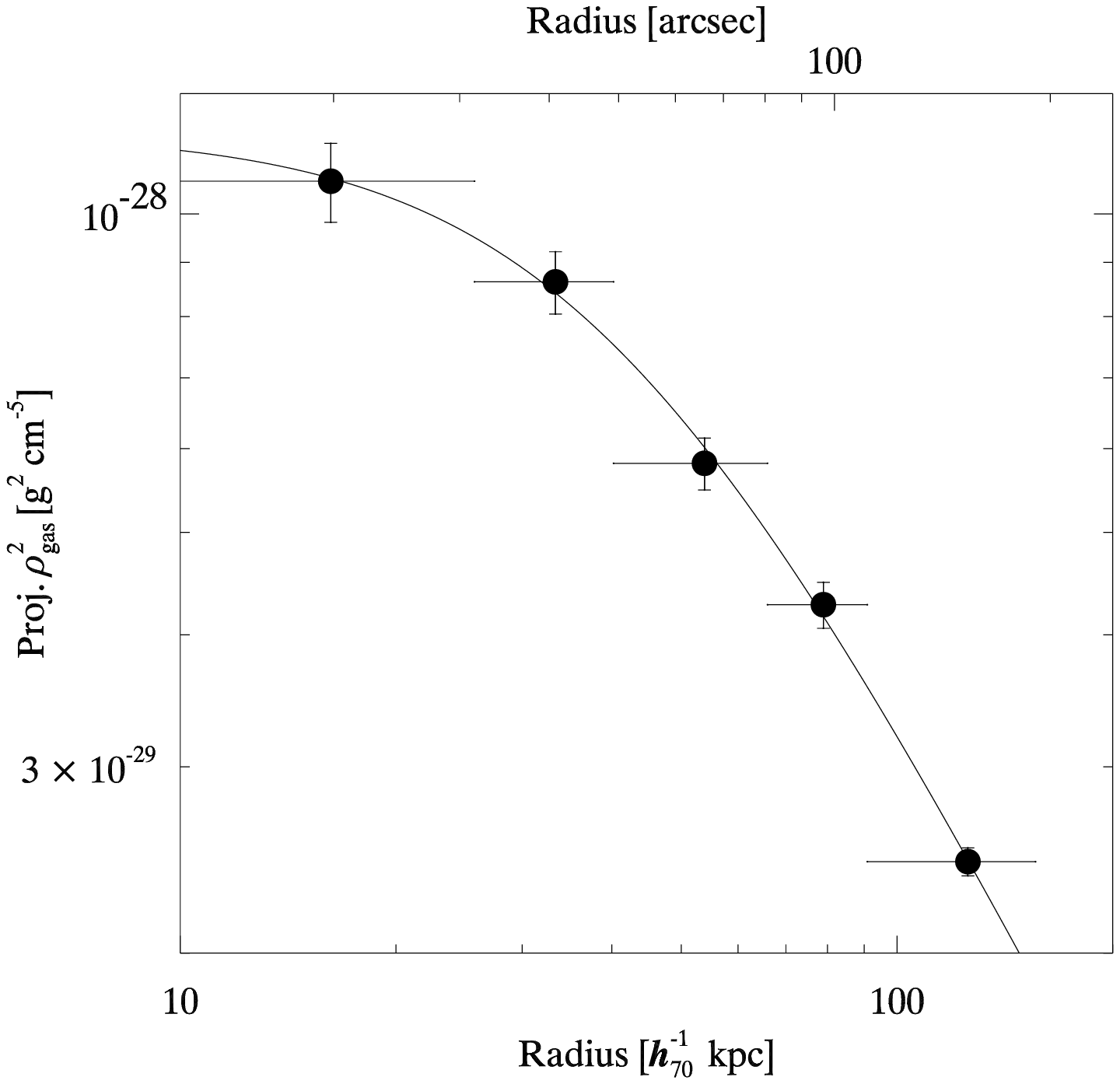,height=0.38\textheight}}
}
\parbox{0.49\textwidth}{
\centerline{\psfig{file=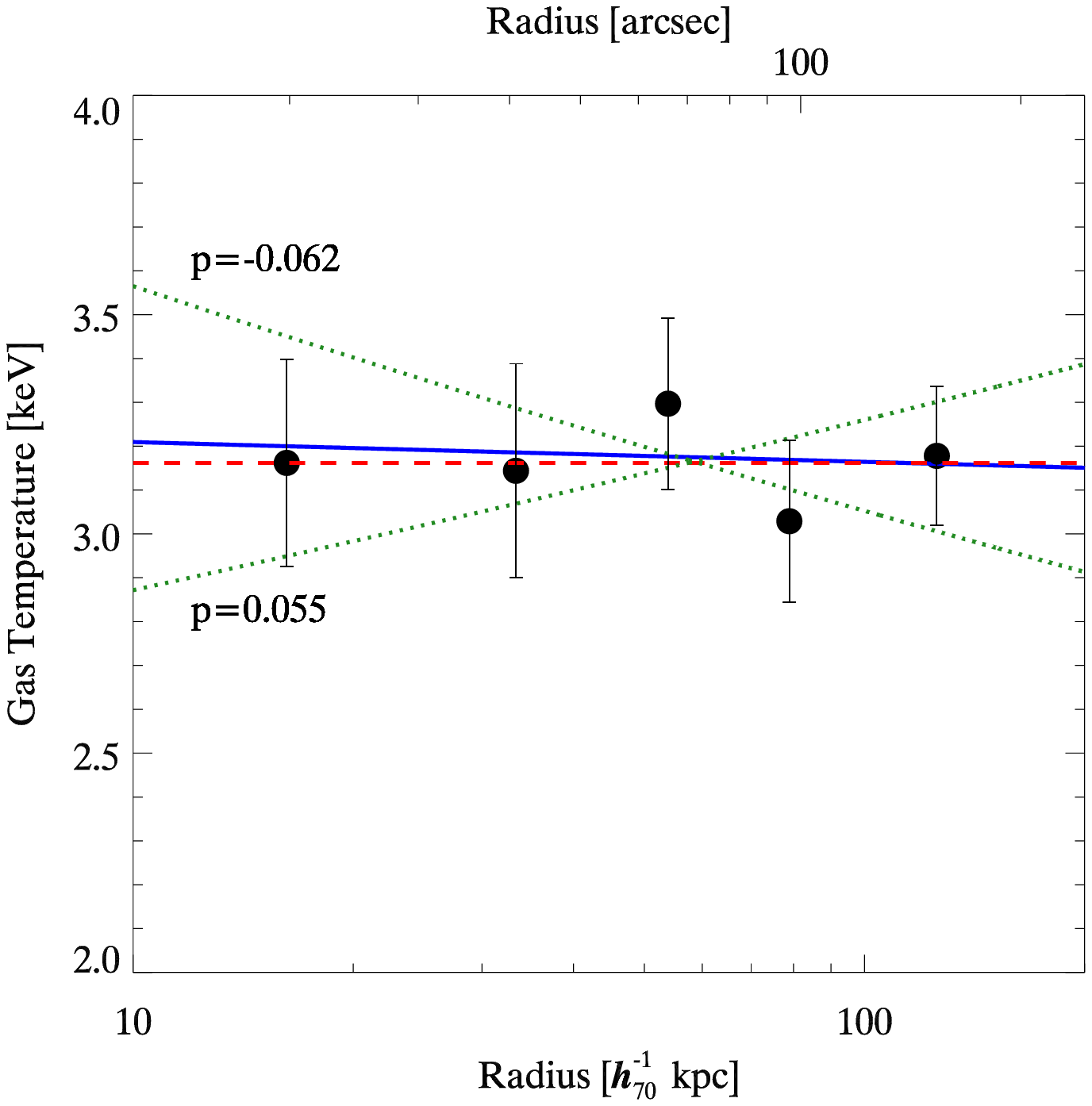,height=0.38\textheight}}
}
\caption{\label{fig.data} {\em Left Panel}: \chandra\ radial profile of 
the projected gas density squared ($\int \rhog^2 dl$) obtained by
dividing the $norm$ parameter of the \apec\ model (Table
\ref{tab.spec}) by the area of the annulus as described in \S
\ref{data}. Horizontal bars indicate the sizes of the annuli used to
extract spectra, and the sizes of the spherical shells used in our
deprojected analysis. Overlaid is the best-fitting $\beta$-model ({\em
solid curve}). {\em Right Panel}: \chandra\ radial temperature profile
of A2589.  Overlaid are the best-fitting isothermal ({\em dashed (red)
curve}), and power-law ({\em solid (blue) curve)} models. Also shown
are the $1\sigma$ limits for the power-law model ({\em dotted (green)
curves)}. The temperature data points refer to projected quantities
(Table \ref{tab.spec}) while the temperature models are deprojected
(Table \ref{tab.gastemp}).}
\end{figure*}

\begin{table*}[t] \footnotesize
\begin{center}
\caption{Deprojected Radial Profiles of Gas Density and Temperature}
\label{tab.gastemp}
\begin{tabular}{lcccc} \tableline\tableline\\[-7pt]
\\[-1pt]

\tableline\\[-5pt]
\multicolumn{5}{c}{Gas Density}\\[+2pt]
\tableline
\\[-7pt]
		& & $r_{c}$ & & $\rhog_0$ \\ Model & $(\chi^2$/dof) &
		(kpc) & $\beta$ & ($10^{-26}$ g cm$^{-3}$)\\
\\[-7pt]
\tableline\\[-5pt]
$\beta$	  	& 0.8/2	 & $40\pm 12$ & $0.39\pm 0.04$	& $1.93\pm 0.12$\\
\tableline\\[-7pt]
\\[-1pt]
\\

\tableline\\[-5pt]
\multicolumn{5}{c}{Temperature}\\[+2pt]
\tableline
\\[-7pt]
		&		 & $T_{100}$ 	&         \\
Model		& $(\chi^2$/dof) & (keV) 	& $p$     \\
\tableline
isothermal	& 1.0/4	 	& $3.16\pm0.22$ & $0$ \\
power law	& 1.0/3	 	& $3.17\pm0.43$ & $-0.006\pm 0.06$ \\

\tableline \\[-1.0cm]
\end{tabular}
\end{center}
\tablecomments{We find a central electron density of
$n_{e_0} = 9.8\pm 0.6\times10^{-3}$ cm$^{-3}$ (the conversion factor
between $n_{e_0}$ and $\rhog_0$ is $5.09\times 10^{23})$. $T_{100}$ is
the temperature of the isothermal model ($p=0$), and is the
temperature evaluated at $r=100$~kpc for the power law model. }
\end{table*}

We first attempted to obtain three-dimensional gas parameters by
performing a spectral deprojection procedure based on the
``onion-peeling'' method that we have used in previous studies
\citep{buot00c,buot03a,lewi03a}. Because the S/N of the A2589 data is
relatively low, we are unable to obtain precise constraints on the
deprojected temperature and abundance profiles using this procedure
(but see \S \ref{sys}). Consequently, we projected parameterized
models of the three-dimensional quantities, \rhog\ and $T$, and fitted
these projected models to the results obtained from our analysis of
the data projected on the sky (Table \ref{tab.spec}). In this manner
we obtained good constraints on the three-dimensional radial profiles
of \rhog\ and $T$. Note that for each annulus we assign a single
radius value, $r \equiv [(r_{\rm out}^{3/2} + r_{\rm
in}^{3/2})/2]^{2/3}$, where $r_{\rm in}$ and $r_{\rm out}$ are
respectively the inner and outer radii of the annulus/shell. We found
this prescription to provide an excellent estimate of the mean
weighted radius in our previous study of A2029 \citep{lewi03a}.

The $norm$ parameter of the \apec\ model obtained from the spectral
fits (see Table \ref{tab.spec}) is proportional to $\int \rhog^2 dV$,
where $dV$ represents the emitting volume. For our data in projection
we have $dV=Adl$, where $A$ is the area of the annulus on the sky and
$dl$ is the length element along the line of sight; i.e., $\int
\rhog^2 dl \propto norm/A$. (It is assumed that the plasma emissivity
is constant along the line of sight which is a very good approximation
for the nearly isothermal spectrum of A2589.) In Figure \ref{fig.data}
we plot $\int
\rhog^2 dl$ along with the best-fitting $\beta$ model
\citep{beta}, 
\begin{eqnarray}
\rhog & = & \rhog_{0}(1 + r^2/r_c^2)^{-3\beta/2} \label{eqn.betamod} \\
\int  \rhog^2 dl & =
&{\Gamma(\frac{1}{2})\Gamma(3\beta-\frac{1}{2}) \over \Gamma(3\beta)} {\rhog_0^2 r_c \over (1 + R^2/r_c^2)^{3\beta-0.5}},
\end{eqnarray}
where the integral is evaluated along the line of sight
$(-\infty,+\infty)$, and $R$ is the radius of the annulus on the sky..
The best-fitting parameters, $1\sigma$ errors, and $\chi^2$ values are
listed in Table \ref{tab.gastemp}. The errors are calculated by
fitting the $norm$ profiles obtained for each Monte Carlo error
simulation (\S \ref{spec}). We find that the simple $\beta$ model is
an excellent fit.

We modeled the temperature data as the emission-weighted projection of
a power law profile,
\begin{eqnarray}
T(r) & = & T_{100}\left(\frac{r}{100 \, {\rm  kpc}}\right)^p\\
\langle T(R) \rangle & = & \int \rhog^2\,
T_{100}\left(\frac{\sqrt{R^2+l^2}}{100 \, {\rm  kpc}}\right)^p dl / \int \rhog^2  dl,
\end{eqnarray}
where $T_{100}$ is the temperature at $r=100$~kpc, \rhog\ is the
$\beta$ model given above, and again $R$ is the radius of the annulus
on the sky. The best-fitting model ($\langle T(R)
\rangle$) is plotted in Figure \ref{fig.data} and the parameters and
$1\sigma$ errors for $T(r)$ are given in Table \ref{tab.gastemp}. The
temperature profile is nearly isothermal. This is illustrated by the
excellent fit obtained for $p$ fixed to 0 (Figure \ref{fig.data} and
Table \ref{tab.gastemp}). 

\subsection{Three-Dimensional Radial Mass Profile}
\label{mass}

\begin{table*}[t] \footnotesize
\begin{center}
\caption{Parameters of Gravitating Matter and Dark Matter Radial Profiles}
\label{tab.mass}
\begin{tabular}{lcccccc} \tableline\tableline\\[-7pt]

		& & & $M_{100}$ & $r_s$ & & \rvir \\ Mass Model &
		$(\chi^2$/dof) & $\gamma$ & $(10^{13}\msun)$ & (kpc) &
		$c$ & (Mpc) \\
\tableline\\[-1pt]

\tableline\\[-5pt]
\multicolumn{7}{c}{Gravitating Matter}\\[+2pt]
\tableline
\\[-7pt]
power law	& 5.0/3	 & $1.63\pm0.14$& $1.11\pm 0.44$ 	& \nodata 	& \nodata 	& \nodata \\
NFW		& 2.0/3	 & \nodata	& \nodata		& $153\pm 362$   & $6.6\pm 2.5$ 	& $1.01\pm 0.32$ \\
\tableline\\[-7pt]
\\[-1pt]
\\

\tableline\\[-5pt]
\multicolumn{7}{c}{Dark Matter}\\[+2pt]
\tableline
\\[-7pt]
power law	& 2.7/2	 & $1.65\pm 0.21$ & $0.90\pm 0.33$ 	& \nodata 	& \nodata 	& \nodata \\
NFW		& 1.9/2	 & \nodata	& \nodata		& $208\pm 865$   & $4.9\pm 2.4$ 	& $1.03\pm 0.51$ \\

\tableline \\[-5pt]

\end{tabular}
\end{center}
\tablecomments{The power-law temperature model is used for all the
fits. The fits to the DM profile exclude the inner mass point
and assume $\mstars/L_V=6.1$ (\S \ref{stars}).}
\end{table*}

\begin{figure*}[ht]
\parbox{0.49\textwidth}{
\centerline{\psfig{file=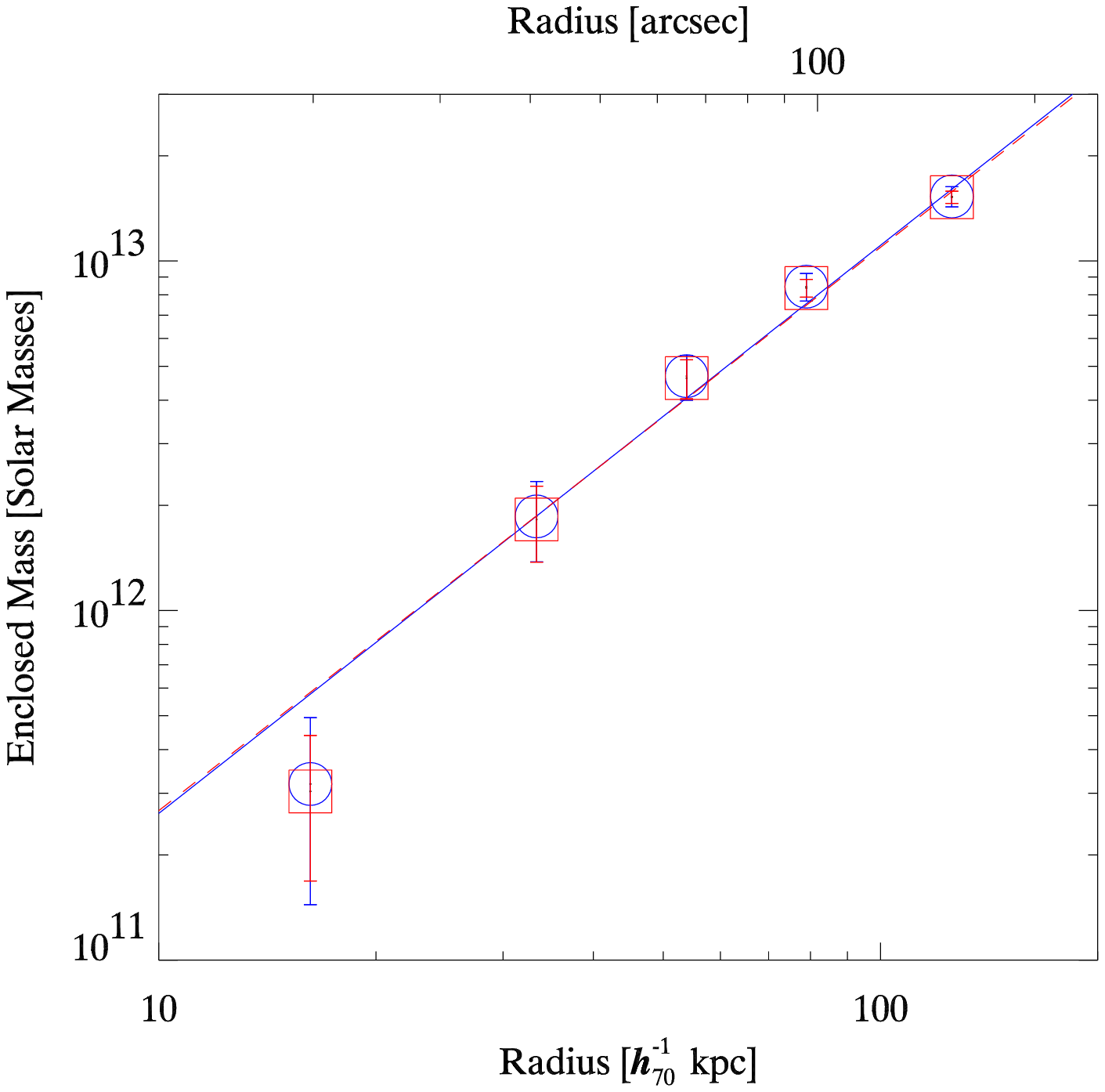,height=0.38\textheight}}
}
\parbox{0.49\textwidth}{
\centerline{\psfig{file=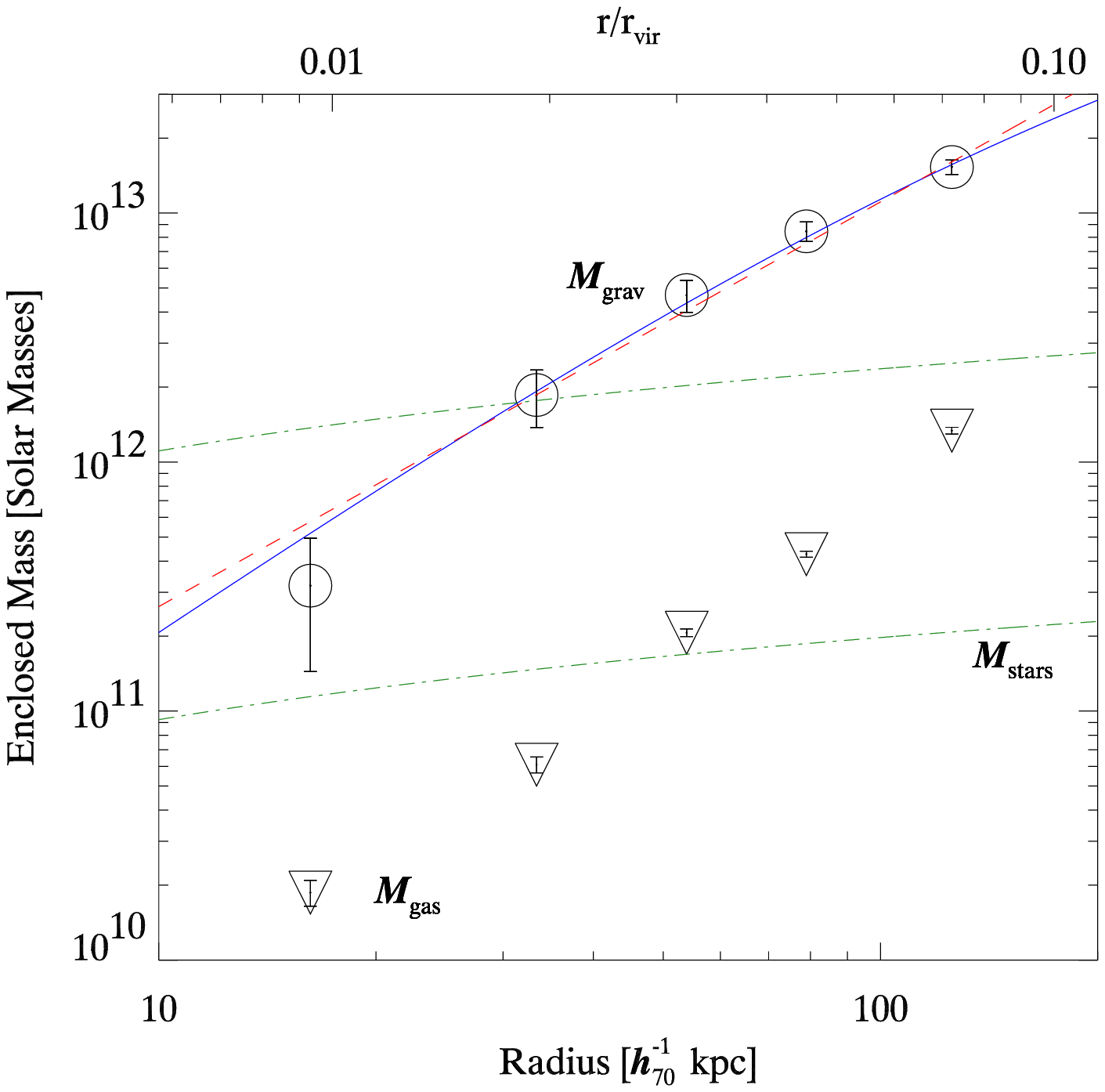,height=0.38\textheight}}
}
\caption{\label{fig.grav} \emph{Left Panel}: Total gravitating mass of
the cluster for each radial bin obtained from the isothermal
(\emph{open squares, red}) and the power-law (\emph{open circles,
blue}) fits to the gas temperatures. The $\beta$-model
parameterization for \rhog\ was used in both cases. The mass points
are fitted with corresponding power law functions (\emph{solid line,
blue}: power-law $T$ model, \emph{dashed line,red}: isothermal $T$
model).  \emph{Right Panel}: Total gravitating mass (data points
enclosed with open circles), overlaid with two different fitted
models: NFW (\emph{(blue) solid curve}) and power-law (\emph{(red)
dashed line}). Each model uses the power-law parameterization of the
temperature profile. The enclosed gas mass is plotted as data points
enclosed with open triangles. Also plotted is the mass range expected
from the stars \emph{((green) dot-dashed curves)}: The lower curve
assumes $\mstars/L_V = 1\msun/\lsun$ and the upper curve $\mstars/L_V
= 12\msun/\lsun$ (see \S \ref{stars}). We have used large open symbols
to identify the data points, as some of the error bars are difficult
to see in this logarithmic plot. The upper axis shows the radius in
units of the virial radius, $\rvir=1.72$~Mpc.}
\end{figure*}

Using the parameterized functions for the gas density and temperature
just described, the radial mass distribution (eq.\ \ref{eqn.he})
becomes,
\begin{equation}
\mgrav (<r) = r{k_BT_{100}(r/100 \, {\rm  kpc})^p \over G\mu m_{\rm p}}
\left( { 3\beta r^2/r_c^2 \over 1 + r^2/r_c^2} - p \right).
\label{eqn.grav} 
\end{equation}
In Figure \ref{fig.grav} we plot \mgrav\ evaluated for each radial bin
obtained from the isothermal (\emph{open squares, red}) and the
power-law (\emph{open circles, blue}) fits to the gas
temperatures.  The mass profiles obtained from the isothermal and
power-law models agree extremely well. Fitting a power law,
\begin{equation}
M(<r) =  M_{100}\left(\frac{r}{100 \, {\rm  kpc}}\right)^{\gamma},
\end{equation}
to the mass profile itself gives $\gamma = 1.63\pm 0.14$ and $\gamma =
1.61\pm 0.13$ for the power-law ($p$) and isothermal ($p=0$)
temperature parameterizations respectively.  The excellent agreement
demonstrates that the mass profile is not particularly sensitive to
the temperature parameterization, and thus we focus on the power-law
temperature parameterization henceforth. 

The NFW density and enclosed mass profile are given by,
\begin{eqnarray}
\rho(r)&  = & \frac{\rho_c(z)\delta_c}{(r/r_s)(1+r/r_s)^2},\\
M(<r) & = & 4\pi \rho_c(z)\delta_c r_s^3\left[\ln \left(\frac{r_s+
r}{r_s}\right) - \frac{r}{r_s + r}\right], \label{eqn.nfw}
\end{eqnarray}
where $\rho_c(z) = 3H(z)^2/8\pi G$ is the critical density of the
universe at redshift $z$, $r_s$ is the scale radius, and $\delta_c$ is
a characteristic dimensionless density which, when expressed in terms
of the concentration parameter $c$, takes the form,
\begin{equation}
\delta_c = \frac{200}{3}\frac{c^3}{\ln(1+c) - c/(1+c)}. \label{eqn.conc}
\end{equation}
Hence, there are two free parameters for the NFW mass profile: $r_s$
and $c$. The virial radius, defined to be the radius where the
enclosed average mass density equals $200\rho_c(z)$, is simply
$\rvir=cr_s$. We calculate $\rho_c(z)$ at the redshift of A2589
assuming $\omegam=0.3$ and $\omegalambda =0.7$ today.

In Figure \ref{fig.grav} we show the result of fitting an NFW model to
the mass profile along with the power-law fit for comparison.  The
parameters for both models are listed in Table \ref{tab.mass}. The NFW
model provides a very good fit to mass points, better than the power
law model, though the power law is still a fair representation of the
mass profile. Both models lie above the innermost data
point. Excluding the innermost data point does not significantly
change the results. For example, when omitting the inner mass point
the power-law mass fit gives $\gamma=1.47\pm 0.15$ which is within
$\approx 1\sigma$ of the value obtained when fitting all the points
(Table \ref{tab.mass}). Similarly, we obtain a concentration parameter
$c=7.8\pm 3.4$ for NFW when excluding the inner point which is quite
consistent with the value given in Table \ref{tab.mass}.  (We mention
that a Hernquist model \citep{hern90} fits as well as NFW
($\chi^2=1.9$ for 3 dof), but the Moore model is not quite as good a
fit ($\chi^2=6.7$ for 3 dof).)

The virial radius (\rvir) obtained from the NFW fit to \mgrav\ is
rather small ($1.01\pm 0.32$~Mpc) for a massive galaxy cluster, though
our value does have a large uncertainty. It is consistent with the
virial radius, $\rvir = 1.72\pm 0.32$~Mpc, obtained using the virial
mass--temperature relation of \citet{math01a} with the temperature of
the isothermal model (Table \ref{tab.spec}).

\section{Gas Mass, Gas Fraction, and $\Omega_{m}$}
\label{gas}

The radial profile of the enclosed mass of hot gas obtained from the
$\beta$-model (eq. \ref{eqn.betamod}) is plotted in Figure
\ref{fig.grav}. The gas fraction $(\gasfrac = \mgas / \mgrav)$ is
$0.059\pm 0.033$ in the inner radial bin and rises to $0.087\pm 0.006$
in the last radial bin; i.e., the gas mass is a minor contributor to
the gravitating mass over the region studied. The gas fraction in the
last radial bin may be safely considered to be a lower limit since
simulations and other measurements at larger radii in clusters suggest
$\gasfrac= (0.2-0.3) h_{70}^{-3/2}$ \citep[e.g.,][]{alle02a}. Assuming
the global baryon fraction in A2589 is representative, then we place
an upper limit on the present matter density, $\omegam \le \Omega_{\rm
b}/\gasfrac = 0.56 \pm 0.05$ where we have used $ \Omega_{\rm
b}h^2=0.024\pm 0.001$ for $h=0.7$ from recent measurements of the CMB
\citep{sper03a}. We emphasize that our upper limit on $\omegam$ results from
underestimating the baryon fraction because (1) we do not measure a
global value of \gasfrac, and (2) we account for only the baryons
associated with the hot gas.

\section{Gravitating Mass-to-Stellar Light Ratio and the Stellar Mass
Contribution}
\label{stars}

\begin{figure*}[t]
\centerline{\psfig{file=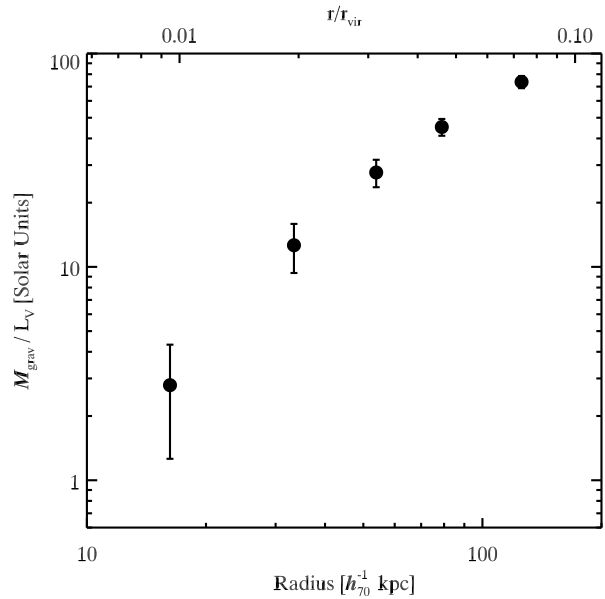,height=0.38\textheight}}
\caption{\label{fig.ml} Ratio of gravitating mass to
stellar light from the cD galaxy in A2589.}
\end{figure*}

To calculate the ratio of gravitating mass to stellar light we used
the fits to the $V$-band surface brightness profile of A2589 published
by \citet{malm85a}. These authors found that a King model with core
radius $r_c=0.98h_{70}^{-1}$~kpc and a total luminosity of $2.3\times
10^{11} L_{V,\sun} h_{70}^{-2}$ provides a a good fit to surface
brightness profile out to $R=130h_{70}^{-1}$~kpc. The King model fit
translates to a volume light density profile, $\rho_V(r) = \rho_V(0) (1
+ r^2/r^2_c)^{-3/2}$, where $\rho_V(0) = 3.8L_{V,\sun} {\rm
pc}^{-3}$. The enclosed luminosity as a function of radius in three
dimensions $(L_V(<r))$ is calculated by integrating $\rho_V$ over the
volume of the sphere of radius $r$. We plot $\mgrav(<r)/L_V(<r)$ in
Figure \ref{fig.ml}.

$\mgrav/L_V$ climbs from $2.8\pm 1.5\, \msun/\lsun$ at the inner
radial bin to $73.6\pm 4.8\, \msun/\lsun$ at the outer radial bin.  At
the resolution of our chosen radial bin sizes there is no evidence for
a sharp transition in $\mgrav/L_V$ over the region studied. 

Converting $L_V$ to stellar mass (\mstars) is problematic because of
the uncertain stellar mass-to-light ratio ($\mstars/L_V$) of
elliptical galaxies. Often stellar dynamical measurements of the cores
of elliptical galaxies are taken as estimates for the stellar
mass. But since such dynamical studies (like X-rays) are sensitive to
the total gravitating mass, they cannot be relied upon to yield robust
values of the stellar mass separated from the DM. Stellar population
synthesis studies offer a direct method to calculate $\mstars/L_V$ but
unfortunately allow for a large range of values ($\mstars/L_V= 1-12\,
\msun/\lsun$) because of uncertainties associated with the stellar
initial mass function, the age of the population(s), and the
metallicity \citep[e.g.,][]{pick85a,mara99a} -- though see
\citet{math89a} for arguments that the stellar mass is similar to the
value determined from stellar dynamics.

In Figure \ref{fig.grav} we plot $\mstars$ for the cases $\mstars/L_V= 1\,
\msun/\lsun$ and  $\mstars/L_V= 12\, \msun/\lsun$ to show the allowed
range from population synthesis studies of elliptical galaxies.  The
upper limit for \mstars\ essentially passes through the value of
\mgrav\ at the second radial bin ($r=33$~kpc) but greatly exceeds
\mgrav\ in the central radial bin. If  \mgrav\  in the central
bin is reliable then $\mstars/L_V$ must be small enough to match
$\mgrav/L_V = 2.8\pm 1.5\, \msun/\lsun$ obtained from the X-ray
analysis. Hence, we assume a plausible range of stellar mass-to-light
ratios, $\mstars/L_V = (1-7.4) \msun/\lsun$, where the upper limit is
the $3\sigma$ upper limit on $\mgrav/L_V$ in the central radial bin
and the lower limit is determined by the population synthesis models.

\section{The Dark Matter Radial Profile}
\label{dm}

\begin{figure*}[t]
\centerline{\psfig{file=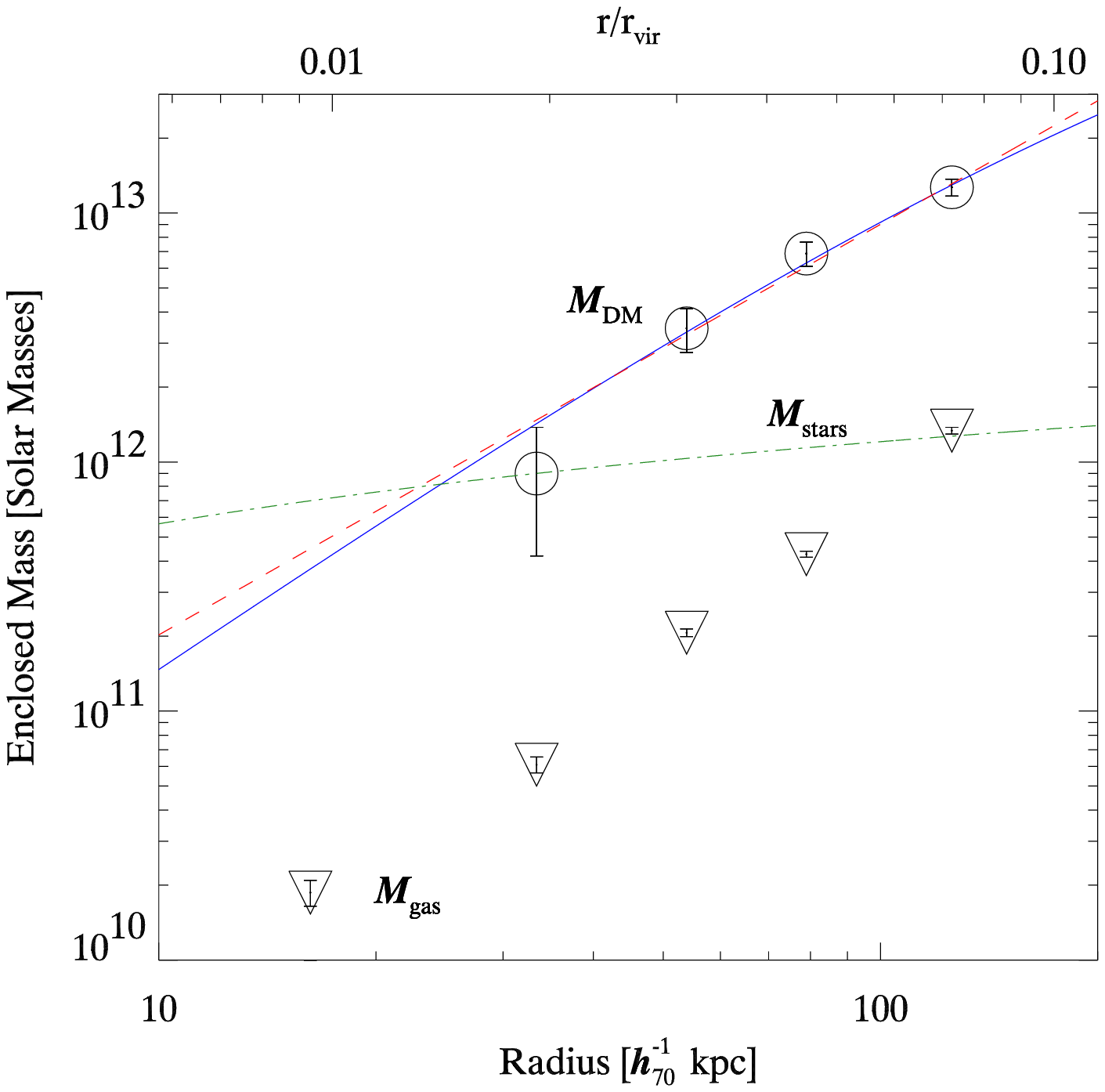,height=0.38\textheight}}
\caption{\label{fig.dm} Same as Figure \ref{fig.grav} except that
\mgrav\ is replaced by \mdm, and \mstars\ is calculated for
$\mstars/L_V = 6.1\, \msun/\lsun$.}
\end{figure*}

Inspection of Figure \ref{fig.grav} reveals that the gravitating mass
exceeds the combined masses of stars and hot gas for $r\ga 0.02\rvir$
indicating that DM dominates over most of the region studied. We write
the mass in DM as, $\mdm = \mgrav - \mgas - \mstars$, where \mgas\ and
\mstars\ are described in \S \ref{gas} and \S \ref{stars}. For
\mstars\ we define a fiducial value by requiring that interior to the
effective radius ($R_e=32.6$~kpc, \citealt{malm85a}) we have
$\mdm=\mstars$. This definition allows for a fairly smooth transition
between stellar-dominated and DM-dominated regimes and yields
$\mstars/L_V = 6.1\, \msun/\lsun$ consistent with the values
determined in \S \ref{stars}.

In Figure \ref{fig.dm} we plot \mstars\ and \mdm\ corresponding to
this fiducial stellar model. The inner data point cannot be seen in
the figure because the dark mass is negative there, $\mdm = -4.0\pm
1.7 \times 10^{11} \msun$, though it has large error. Consequently, we
exclude the inner radial bin from fits to \mdm\ when using the fiducial
value of $\mstars/L_V$.

We plot the best-fitting power-law and NFW models to \mdm\ in Figure
\ref{fig.dm}; the fit parameters are listed in Table \ref{tab.mass}.
The quality of the fits and the parameter values of the DM models are
consistent with those obtained for the gravitating mass within the
$1\sigma$ errors. (Note: including the inner (negative) data point
results in fits that are of significantly lower quality: $\chi^2=19.4$
for power law and $\chi^2=18.5$ for NFW, each with 3 dof.) Similar to
the fits to \mgrav, the Moore model does not fit quite as well as NFW
($\chi^2=3.6$ for 2 dof), but the Hernquist model provides a
comparable fit ($\chi^2=1.8$ for 2 dof). The core parameter for the
Hernquist model is poorly constrained ($a=0.34\pm 0.77$~Mpc).

\section{Systematic Error Budget for \mdm}
\label{sys}

\begin{table*}[t] \footnotesize
\begin{center}
\caption{Error Budget for Dark Matter Fits}
\label{tab.sys}
\begin{tabular}{ccccccc}  \tableline\tableline\\[-7pt]
& Best & $\Delta$Statistical & $\Delta$Background & $\Delta$\mstars &
$\Delta$Deproj\\
\tableline \\[-7pt]
Power Law:\\
$\gamma$  & $1.65$ & $\pm 0.21$ & $(+0.17,-0.16)$ & $(+0.06,-0.00)$ & $+0.35$\\ 
\\
NFW:\\
$c$ & 4.9 & $\pm 2.4$ & $(+1.9,-1.7)$ & $(+1.2,-0.6)$ & $-3.8$\\
\rvir (Mpc) & 1.0 & $\pm 0.5$ & $(+0.3,-0.2)$ & $(+0.06,-0.05)$ & $+1.9$\\
\tableline \\[-7pt]
\tableline \\[-1.0cm]
\end{tabular}
\tablecomments{The ``Best'' column indicates the best-fit value and
``$\Delta$Statistical'' the $1\sigma$ statistical error from Table
\ref{tab.mass}. ``$\Delta$Background'' gives the results when the
X-ray background level is set to $\pm 20\%$ of
nominal. ``$\Delta$\mstars'' represents the uncertainty associated
with the stellar mass-to-light ratio. ``$\Delta$Deproj'' provides an
estimate of the error associated with the deprojection procedure.}
\end{center}
\end{table*}

This section provides an assessment of the magnitude of systematic
errors on the fits to \mdm.

\smallskip
\noindent{\it X-ray Background:} To examine the sensitivity of the measured
dark matter properties to reasonable background errors we repeated our
entire analysis using the standard background templates and correction
files (\S \ref{obs}) renormalized to have count rates $\pm 20\%$ of
their nominal values. We find that the derived gas density profile is
hardly affected while the temperature profile is modified within the
$1\sigma-1.5\sigma$ errors: $p=(-0.09,0.07)$ for these different
background levels. The effect on the DM fits is less than the
$1\sigma$ errors in the DM model parameters as shown in Table
\ref{tab.sys}. We also examined whether the DM fits were
systematically different if the normalization of the background was
allowed to be a free parameter during the fits. We found that the DM
fits are fully consistent with the results quoted above. We note,
however, that in this case the best-fitting temperatures ($\approx
20\%$) and abundances ($\approx 40\%$) are lower than the values
listed in Table \ref{tab.spec}, although the differences are
significant only at the $\approx 2\sigma$ level.

\smallskip
\noindent{\it Stellar Mass:}  In Table \ref{tab.sys} we quote
the ranges in model parameters fitted to \mdm\ considering the
variation in \mstars\ implied by $\mstars/L_V = (1-7.4)\, \msun/\lsun$
(see \S \ref{stars}).  Note that for $\mstars/L_V = 1
\msun/\lsun$ the inner radial bin for
\mdm\ is positive, and so we included it in the fits in Table
\ref{tab.sys}. Uncertainties associated with the stellar mass are
within the $1\sigma$ errors.

\smallskip
\noindent{\it Center:} Since the X-ray centroid varies with radius we
examined the sensitivity of our results to the centroid. For
comparison to our results obtained using the centroid computed within
a circular aperture of radius $20\arcsec$, we used a radius of
$3\arcmin$. Using this larger radius we obtain a centroid offset by
$\approx 8\arcsec$ from the previous value. However, we find that the
results obtained for the DM fits change negligibly ($<1\%$) when using
the different centroid.

\smallskip
\noindent{\it Deprojection Method:} Our method to deproject the gas
density and temperature uses the projections of parameterized models
(\S \ref{data}). For comparison, we also performed a non-parametric
deprojection using the ``onion-peeling'' method as implemented in our
previous studies \citep[e.g.,][]{buot00c,buot03a,lewi03a}. To obtain
stable deprojected parameters for the A2589 \chandra\ data we imposed
the following restrictions: (1) the deprojected temperatures in annuli
\#3-5 were confined to the $1\sigma$ limits in Table
\ref{tab.spec}, and (2) the deprojected  metallicities were set to
those in Table \ref{tab.spec}. Because of these restrictions {\sl the
results obtained from this deprojection should only be viewed as an
indicator of the sensitivity of the results to deprojection and not as
the preferred values.} We found that the deprojected temperature and
density profiles give results consistent with those obtained from the
parametric deprojection when fitted with the same models in three
dimensions. However, the gas density is more centrally peaked than
indicated by the $\beta$ model and thus, like in A2029
\citep{lewi03a}, we find that adding a central cusp to the $\beta$
model provides a better fit. We find that \mdm\ is small, but
positive, in the central bin. Using all five data points we obtain the
results reflected in Table \ref{tab.sys} for fits to the DM.  

\section{Conclusions}
\label{conc}

Our analysis of the \chandra\ data indicates that from the largest
radius probed ($r=0.07\rvir$) down to $r\approx 0.02\rvir$ the
dominant contributor to the gravitating matter in A2589 is the
DM. Over this region the radial profile of the DM is fitted well by
the NFW and Hernquist profiles predicted by CDM. (These models fit the
gravitating matter well down to $r=0.009\rvir$.)  The inferred value
of \rvir\ of the NFW model is consistent (within the relatively large
measurement errors) with that expected from CDM simulations (1.7~Mpc,
see \S \ref{mass}); equivalently, the concentration parameter
$c=4.9\pm 2.4$ is in the range expected for a massive cluster
\citep{nfw}. The DM profile over this region is also described
well by a power law, $M(<r)\propto r^{\gamma}\rightarrow \rho\propto
r^{-\alpha}$ where $\alpha = 3 - \gamma$. We find $\alpha=1.35\pm
0.21$ (statistical error, see \S \ref{sys}) which is consistent with
the NFW value $\alpha=1$ (and Moore value of $\alpha=1.5$) but is
significantly larger than $\alpha\approx 0.5$ found in LSB galaxies
\citep[e.g.,][]{swat00a} and expected from SIDM models \citep{sper00}.
We note that the profile of the gravitating matter ($\alpha=1.37\pm
0.14$) is also very consistent with the DM profile and with CDM
predictions \citep{elza03a}.

The DM properties of A2589 agree very well with those obtained for
A2029 \citep{lewi03a}. Because the cores of these clusters are
undisturbed by interactions from a central radio source, the results
we have obtained for their core DM distributions provide critical
confirmation of (in some respects) similar results obtained from other
otherwise-relaxed clusters with disturbed cores (most notably Hydra-A,
\citealt{davi01a}) and clarify ambiguous cases like A1795
\citep{etto02a}. Furthermore, the evidence that the NFW-CDM profile is
a good fit to cluster DM halos on larger scales $(0.1\rvir\la r<
0.5\rvir)$ \citep[e.g.,][]{alle02a,prat02a} is extended down to $r\sim
0.01\rvir$ by our complementary analyses of the \chandra\ data of
A2029 and A2589.  This agreement of cluster DM profiles with the NFW
model is also supported by the recent weak-lensing analysis of several
clusters by \citet{dahl03a} and the joint lensing-X-ray study of
\citet{arab02a}.

In sum, the current evidence from X-ray and lensing studies indicates
that the radial DM profiles for cluster-sized DM halos are consistent
with CDM predictions within the expected cosmological scatter
\citep{bull01a}. The disagreement between the NFW-CDM profile and the
flatter profiles observed for LSB galaxies indicates that either CDM
simulations are adequate for clusters but inadequate for LSB galaxies
or the interpretations of the measurements of the DM profiles for LSB
galaxies are in error. Some recent studies have indeed suggested that
systematic errors could account for the flat density profiles obtained
for LSB galaxies \citep[e.g.,][]{swat03a}.

Future X-ray and lensing studies of the DM in other (relaxed) cluster
cores are essential to refine the constraints on the DM properties
obtained from these initial studies and to quantify precisely the
cosmological scatter.

\acknowledgments

Partial support for this work was provided by the National Aeronautics
and Space Administration through Chandra Award Number GO2-3170X issued
by the Chandra X-ray Observatory Center, which is operated by the
Smithsonian Astrophysical Observatory for and on behalf of the
National Aeronautics and Space Administration under contract
NAS8-39073.

\bibliographystyle{apj}

\begin{thebibliography}{46}
\expandafter\ifx\csname natexlab\endcsname\relax\def\natexlab#1{#1}\fi

\bibitem[{{Allen} {et~al.}(2002){Allen}, {Schmidt}, \& {Fabian}}]{alle02a}
{Allen}, S.~W., {Schmidt}, R.~W., \& {Fabian}, A.~C. 2002, \mnras, 334, L11

\bibitem[{{Arabadjis} {et~al.}(2002){Arabadjis}, {Bautz}, \&
  {Garmire}}]{arab02a}
{Arabadjis}, J.~S., {Bautz}, M.~W., \& {Garmire}, G.~P. 2002, \apj, 572, 66

\bibitem[{{Arnaud}(1996)}]{xspec}
{Arnaud}, K.~A. 1996, in ASP Conf. Ser. 101: Astronomical Data Analysis
  Software and Systems V, Vol.~5, 17

\bibitem[{{Bauer} {et~al.}(2000){Bauer}, {Condon}, {Thuan}, \&
  {Broderick}}]{baue00a}
{Bauer}, F.~E., {Condon}, J.~J., {Thuan}, T.~X., \& {Broderick}, J.~J. 2000,
  \apjs, 129, 547

\bibitem[{{Brighenti} \& {Mathews}(2003)}]{brig03a}
{Brighenti}, F. \& {Mathews}, W.~G. 2003, \apj, 587, 580

\bibitem[{{Bullock} {et~al.}(2001){Bullock}, {Kolatt}, {Sigad}, {Somerville},
  {Kravtsov}, {Klypin}, {Primack}, \& {Dekel}}]{bull01a}
{Bullock}, J.~S., {Kolatt}, T.~S., {Sigad}, Y., {Somerville}, R.~S.,
  {Kravtsov}, A.~V., {Klypin}, A.~A., {Primack}, J.~R., \& {Dekel}, A. 2001,
  \mnras, 321, 559

\bibitem[{{Buote}(2000)}]{buot00c}
{Buote}, D.~A. 2000, \apj, 539, 172

\bibitem[{{Buote} {et~al.}(2003{\natexlab{a}}){Buote}, {Lewis}, {Brighenti}, \&
  {Mathews}}]{buot03a}
{Buote}, D.~A., {Lewis}, A.~D., {Brighenti}, F., \& {Mathews}, W.~G.
  2003{\natexlab{a}}, \apj, 594, 741

\bibitem[{{Buote} {et~al.}(2003{\natexlab{b}}){Buote}, {Lewis}, {Brighenti}, \&
  {Mathews}}]{buot03b}
---. 2003{\natexlab{b}}, \apj, 595, 151

\bibitem[{{Buote} \& {Tsai}(1996)}]{buot96b}
{Buote}, D.~A. \& {Tsai}, J.~C. 1996, \apj, 458, 27

\bibitem[{{Cavaliere} \& {Fusco-Femiano}(1978)}]{beta}
{Cavaliere}, A. \& {Fusco-Femiano}, R. 1978, \aap, 70, 677

\bibitem[{{Dahle} {et~al.}(2003){Dahle}, {Hannestad}, \&
  {Sommer-Larsen}}]{dahl03a}
{Dahle}, H., {Hannestad}, S., \& {Sommer-Larsen}, J. 2003, \apjl, 588, L73

\bibitem[{{Dav{\' e}} {et~al.}(2001){Dav{\' e}}, {Spergel}, {Steinhardt}, \&
  {Wandelt}}]{dave01a}
{Dav{\' e}}, R., {Spergel}, D.~N., {Steinhardt}, P.~J., \& {Wandelt}, B.~D.
  2001, \apj, 547, 574

\bibitem[{{David} {et~al.}(1996){David}, {Jones}, \& {Forman}}]{davi96a}
{David}, L.~P., {Jones}, C., \& {Forman}, W. 1996, \apj, 473, 692

\bibitem[{{David} {et~al.}(2001){David}, {Nulsen}, {McNamara}, {Forman},
  {Jones}, {Ponman}, {Robertson}, \& {Wise}}]{davi01a}
{David}, L.~P., {Nulsen}, P.~E.~J., {McNamara}, B.~R., {Forman}, W., {Jones},
  C., {Ponman}, T., {Robertson}, B., \& {Wise}, M. 2001, \apj, 557, 546

\bibitem[{{Dubinski}(1998)}]{dubi98a}
{Dubinski}, J. 1998, \apj, 502, 141

\bibitem[{{El-Zant} {et~al.}(2003){El-Zant}, {Hoffman}, {Primack}, {Combes}, \&
  {Shlosman}}]{elza03a}
{El-Zant}, A., {Hoffman}, Y., {Primack}, J., {Combes}, F., \& {Shlosman}, I.
  2003, apJ Letters, submitted (astro-ph/0309412)

\bibitem[{{Ettori} {et~al.}(2002){Ettori}, {Fabian}, {Allen}, \&
  {Johnstone}}]{etto02a}
{Ettori}, S., {Fabian}, A.~C., {Allen}, S.~W., \& {Johnstone}, R.~M. 2002,
  \mnras, 331, 635

\bibitem[{{Evrard} {et~al.}(1996){Evrard}, {Metzler}, \& {Navarro}}]{evra96a}
{Evrard}, A.~E., {Metzler}, C.~A., \& {Navarro}, J.~F. 1996, \apj, 469, 494

\bibitem[{{Fabian} {et~al.}(2000){Fabian}, {Sanders}, {Ettori}, {Taylor},
  {Allen}, {Crawford}, {Iwasawa}, {Johnstone}, \& {Ogle}}]{fabi00_perseus}
{Fabian}, A.~C., {Sanders}, J.~S., {Ettori}, S., {Taylor}, G.~B., {Allen},
  S.~W., {Crawford}, C.~S., {Iwasawa}, K., {Johnstone}, R.~M., \& {Ogle}, P.~M.
  2000, \mnras, 318, L65

\bibitem[{{Gastaldello} \& {Molendi}(2002)}]{gast02a}
{Gastaldello}, F. \& {Molendi}, S. 2002, \apj, 572, 160

\bibitem[{{Grevesse} \& {Sauval}(1998)}]{grsa}
{Grevesse}, N. \& {Sauval}, A.~J. 1998, Space Science Reviews, 85, 161

\bibitem[{{Hernquist}(1990)}]{hern90}
{Hernquist}, L. 1990, \apj, 356, 359

\bibitem[{{Jones} \& {Forman}(1999)}]{jone99a}
{Jones}, C. \& {Forman}, W. 1999, \apj, 511, 65

\bibitem[{{Lewis} {et~al.}(2003){Lewis}, {Buote}, \& {Stocke}}]{lewi03a}
{Lewis}, A.~D., {Buote}, D.~A., \& {Stocke}, J.~T. 2003, \apj, 586, 135

\bibitem[{{Lewis} {et~al.}(2002){Lewis}, {Stocke}, \& {Buote}}]{lewi02a}
{Lewis}, A.~D., {Stocke}, J.~T., \& {Buote}, D.~A. 2002, \apjl, 573, L13

\bibitem[{{Malumuth} \& {Kirshner}(1985)}]{malm85a}
{Malumuth}, E.~M. \& {Kirshner}, R.~P. 1985, \apj, 291, 8

\bibitem[{{Maraston}(1999)}]{mara99a}
{Maraston}, C. 1999, in ASP Conf. Ser. 163: Star Formation in Early Type
  Galaxies, 28

\bibitem[{{Markevitch} {et~al.}(2003){Markevitch}, {Bautz}, {Biller}, {Butt},
  {Edgar}, {Gaetz}, {Garmire}, {Grant}, {Green}, {Juda}, {Plucinsky},
  {Schwartz}, {Smith}, {Vikhlinin}, {Virani}, {Wargelin}, \& {Wolk}}]{mark03a}
{Markevitch}, M., {Bautz}, M.~W., {Biller}, B., {Butt}, Y., {Edgar}, R.,
  {Gaetz}, T., {Garmire}, G., {Grant}, C.~E., {Green}, P., {Juda}, M.,
  {Plucinsky}, P.~P., {Schwartz}, D., {Smith}, R., {Vikhlinin}, A., {Virani},
  S., {Wargelin}, B.~J., \& {Wolk}, S. 2003, \apj, 583, 70

\bibitem[{{Mathews}(1989)}]{math89a}
{Mathews}, W.~G. 1989, \aj, 97, 42

\bibitem[{{Mathiesen} {et~al.}(1999){Mathiesen}, {Evrard}, \& {Mohr}}]{math99a}
{Mathiesen}, B., {Evrard}, A.~E., \& {Mohr}, J.~J. 1999, \apjl, 520, L21

\bibitem[{{Mathiesen} \& {Evrard}(2001)}]{math01a}
{Mathiesen}, B.~F. \& {Evrard}, A.~E. 2001, \apj, 546, 100

\bibitem[{{McWilliam}(1997)}]{mcwi97}
{McWilliam}, A. 1997, \araa, 35, 503

\bibitem[{{Mohr} {et~al.}(1995){Mohr}, {Evrard}, {Fabricant}, \&
  {Geller}}]{mohr95a}
{Mohr}, J.~J., {Evrard}, A.~E., {Fabricant}, D.~G., \& {Geller}, M.~J. 1995,
  \apj, 447, 8

\bibitem[{{Moore} {et~al.}(1999){Moore}, {Quinn}, {Governato}, {Stadel}, \&
  {Lake}}]{moor99a}
{Moore}, B., {Quinn}, T., {Governato}, F., {Stadel}, J., \& {Lake}, G. 1999,
  \mnras, 310, 1147

\bibitem[{{Navarro} {et~al.}(1997){Navarro}, {Frenk}, \& {White}}]{nfw}
{Navarro}, J.~F., {Frenk}, C.~S., \& {White}, S.~D.~M. 1997, \apj, 490, 493

\bibitem[{{Perlmutter} {et~al.}(1999){Perlmutter}, {Aldering}, {Goldhaber},
  {Knop}, {Nugent}, {Castro}, {Deustua}, {Fabbro}, {Goobar}, {Groom}, {Hook},
  {Kim}, {Kim}, {Lee}, {Nunes}, {Pain}, {Pennypacker}, {Quimby}, {Lidman},
  {Ellis}, {Irwin}, {McMahon}, {Ruiz-Lapuente}, {Walton}, {Schaefer}, {Boyle},
  {Filippenko}, {Matheson}, {Fruchter}, {Panagia}, {Newberg}, {Couch}, \& {The
  Supernova Cosmology Project}}]{perl99a}
{Perlmutter}, S., {Aldering}, G., {Goldhaber}, G., {Knop}, R.~A., {Nugent}, P.,
  {Castro}, P.~G., {Deustua}, S., {Fabbro}, S., {Goobar}, A., {Groom}, D.~E.,
  {Hook}, I.~M., {Kim}, A.~G., {Kim}, M.~Y., {Lee}, J.~C., {Nunes}, N.~J.,
  {Pain}, R., {Pennypacker}, C.~R., {Quimby}, R., {Lidman}, C., {Ellis}, R.~S.,
  {Irwin}, M., {McMahon}, R.~G., {Ruiz-Lapuente}, P., {Walton}, N., {Schaefer},
  B., {Boyle}, B.~J., {Filippenko}, A.~V., {Matheson}, T., {Fruchter}, A.~S.,
  {Panagia}, N., {Newberg}, H.~J.~M., {Couch}, W.~J., \& {The Supernova
  Cosmology Project}. 1999, \apj, 517, 565

\bibitem[{{Pickles}(1985)}]{pick85a}
{Pickles}, A.~J. 1985, \apj, 296, 340

\bibitem[{{Pratt} \& {Arnaud}(2002)}]{prat02a}
{Pratt}, G.~W. \& {Arnaud}, M. 2002, \aap, 394, 375

\bibitem[{{Sanders} \& {Fabian}(2002)}]{sand02a}
{Sanders}, J.~S. \& {Fabian}, A.~C. 2002, \mnras, 331, 273

\bibitem[{{Spergel} \& {Steinhardt}(2000)}]{sper00}
{Spergel}, D.~N. \& {Steinhardt}, P.~J. 2000, Physical Review Letters, 84, 3760

\bibitem[{{Spergel} {et~al.}(2003){Spergel}, {Verde}, {Peiris}, {Komatsu},
  {Nolta}, {Bennett}, {Halpern}, {Hinshaw}, {Jarosik}, {Kogut}, {Limon},
  {Meyer}, {Page}, {Tucker}, {Weiland}, {Wollack}, \& {Wright}}]{sper03a}
{Spergel}, D.~N., {Verde}, L., {Peiris}, H.~V., {Komatsu}, E., {Nolta}, M.~R.,
  {Bennett}, C.~L., {Halpern}, M., {Hinshaw}, G., {Jarosik}, N., {Kogut}, A.,
  {Limon}, M., {Meyer}, S.~S., {Page}, L., {Tucker}, G.~S., {Weiland}, J.~L.,
  {Wollack}, E., \& {Wright}, E.~L. 2003, apJ, submitted (astro-ph/0302209)

\bibitem[{{Swaters} {et~al.}(2000){Swaters}, {Madore}, \&
  {Trewhella}}]{swat00a}
{Swaters}, R.~A., {Madore}, B.~F., \& {Trewhella}, M. 2000, \apjl, 531, L107

\bibitem[{{Swaters} {et~al.}(2003){Swaters}, {Verheijen}, {Bershady}, \&
  {Andersen}}]{swat03a}
{Swaters}, R.~A., {Verheijen}, M.~A.~W., {Bershady}, M.~A., \& {Andersen},
  D.~R. 2003, \apjl, 587, L19

\bibitem[{{Townsley} {et~al.}(2002){Townsley}, {Broos}, {Nousek}, \&
  {Garmire}}]{town02}
{Townsley}, L.~K., {Broos}, P.~S., {Nousek}, J.~A., \& {Garmire}, G.~P. 2002,
  nuclear Instruments and Methods in Physics Research, in press
  (astro-ph/0111031)

\bibitem[{{Tsai} {et~al.}(1994){Tsai}, {Katz}, \& {Bertschinger}}]{tsai94a}
{Tsai}, J.~C., {Katz}, N., \& {Bertschinger}, E. 1994, \apj, 423, 553

\end{thebibliography}

\end{document}